\documentclass[graybox]{svmult}

\usepackage{mathptmx}       
\usepackage{helvet}         
\usepackage{courier}        
\usepackage{type1cm}        
\usepackage{makeidx}         
\usepackage{graphicx}        
\usepackage{multicol}        
\usepackage[bottom]{footmisc}

\makeindex             

\begin{document}

\title*{Finite-time singularities in modified
  $\mathcal{F}(R,G)$-gravity and singularity
              avoidance}
\author{Lorenzo Sebastiani}
\institute{Lorenzo Sebastiani \at Dipartimento di Fisica, Universit\`a di Trento, Via Sommarive 14, \email{l.sebastiani@science.unitn.it}
}
%
%
\maketitle

\abstract{We study finite-time future singularities in $\mathcal{F}(R,G)$-gravity,
where $R$ and $G$ are the Ricci scalar and the Gauss-Bonnet invariant,
respectively.
In particular, we reconstruct the $F(G)$-gravity and
$\mathcal{F}(R,G)$-gravity models 
realizing the finite-time future singularities.
We discuss a possible way to cure the finite-time future
singularities in $\mathcal{F}(R,G)$-gravity
by taking into account higher-order curvature corrections
or effects of viscous fluids.}


\section{Introduction}
\label{sec:1}
%
Among the possible alternatives in order to explain the Dark Energy Issue, are
the so called Modified Theories of Gravity\cite{f(R)-gravity, F(G)-gravity}.  

We would like to consider modified $\mathcal{F}(R,G)$-gravity, where the action is described by a function of the Ricci scalar $R$ and the Gauss-Bonnet invariant $G=R^{2}-4R_{\mu\nu}R^{\mu\nu}+R_{\mu\nu\xi\sigma}R^{\mu\nu\xi\sigma}$.

Many of modified gravity models 
bring the future universe evolution to finite-time singularities.
Some of these singularities are softer than other and not all physical quantities (scale factor, effective energy density and pressure) necessarly diverge at this finite future time. Note that singular solutions correspond to accelerated universe, and often appear as the final
evolution of unstable de Sitter space.

The presence of finite-time singularities may cause serious problems in the black holes or stellar astrophysics\cite{Maeda}. Thus, it is of some interest to explore the $\mathcal{F}(R,G)$-gravity models realizing singularities and 
if any natural scenario to cure such singularities exists. 


We use units of $k_\mathrm{B} = c = \hbar = 1$ and denote the
gravitational constant $8 \pi G_{N}$ by
${\kappa}^2 \equiv 8\pi/{M_{\mathrm{Pl}}}^2$
with the Planck mass $M_{\mathrm{Pl}} = G_{N}^{-1/2} =
1.2 \times 10^{19}$GeV.

\section{The Model}
\label{sec:2}
The action of $\mathcal{F}(R,G)$-gravity is given by
\begin{eqnarray}
S = \int d^4 x \sqrt{-g} \left[ \frac{\mathcal{F}(R,G)}{2\kappa^2}
+{\mathcal{L}}_{\mathrm{matter}} \right]\,,
\label{azione}
\end{eqnarray}
where $g$ is the determinant of the metric tensor $g_{\mu\nu}$
and ${\mathcal{L}}_{\mathrm{matter}}$ is the matter Lagrangian.
The spatially-flat FRW space-time is described by the metric
\begin{equation}
ds^{2}=-dt^{2}+a^{2}(t)d \mathbf{x}^{2}\,,
\label{metric}
\end{equation}
where $a(t)$ is the scale factor of the universe. 

From the action in Eq.~(\ref{azione}), the FRW-equations of motion (EOM) are derived as 
\begin{equation}
\rho_{\mathrm{eff}}=\frac{3}{\kappa^{2}}H^{2}\,,
\quad
p_{\mathrm{eff}}=-\frac{1}{\kappa^{2}} \left( 2\dot H+3H^{2} \right)\,,
\label{GutenTag}
\end{equation}
where $\rho_{\mathrm{eff}}$ and $p_{\mathrm{eff}}$ are
the effective energy density and pressure of the universe, respectively, and
these are defined as
\begin{eqnarray}
\rho_{\mathrm{eff}} &=&
\frac{1}{{\mathcal{F}}'_{R}} \left\{ \rho +
\frac{1}{2\kappa^{2}}
\left[ \left( {\mathcal{F}}'_{R}R-\mathcal{F} \right)
-6H{\dot{\mathcal{F}}}'_{R}
+G{\mathcal{F}}'_{G}-24H^3{\dot{\mathcal{F}}}'_{G}
\right] \right\}\,,
\label{eq:2.7} \\
p_{\mathrm{eff}} &=&
\frac{1}{{\mathcal{F}}'_{R}} \biggl\{ p +
\frac{1}{2\kappa^{2}} \Bigl[
-\left( {\mathcal{F}}'_{R}R-\mathcal{F} \right)
+4H{\dot{\mathcal{F}}}'_{R}+2{\ddot{\mathcal{F}}}'_{R}
-G{\mathcal{F}}'_{G}\nonumber \\
&& \hspace{10mm}
+16H\left(\dot{H} +H^2 \right){\dot{\mathcal{F}}}'_{G}
+8H^2 {\ddot{\mathcal{F}}}'_{G}
\Bigr]
\biggr\}\,.
\label{eq:2.8}
\end{eqnarray}
Here, $H=\dot{a}(t)/a(t)$ is the Hubble parameter and
the dot denotes the time derivative. $\rho$ and $p$ are the energy density and pressure of matter, whereas 
$
\mathcal{F}'_{R} =
\partial_{R} \mathcal{F}(R,G)$ and
$\mathcal{F}'_{G} =
\partial_G \mathcal{F}(R,G)$. 
For general relativity with $\mathcal{F}(R,G)=R$,
$\rho_{\mathrm{eff}} = \rho$ and $p_{\mathrm{eff}} = p$ and
therefore Eqs.~(\ref{eq:2.7}) and (\ref{eq:2.8}) are the Friedmann equations.
Consequently, Eqs.~(\ref{eq:2.7}) and (\ref{eq:2.8}) imply that
the contribution of modified gravity can formally be included in
the effective energy density and pressure of the universe.

\section{Finite-time future singularities}
\label{sec:3}

We consider the case in which the Hubble parameter is expressed as
\begin{equation}
H=\frac{h}{(t_{0}-t)^{\beta}}+H_{0}\,,
\label{Hsingular}
\end{equation}
where $h$, $t_{0}$ and $H_{0}$ are positive constants and $t<t_{0}$ because it should be for expanding universe.
$\beta$ is a positive constant or a negative non-integer number, so that, when $t$ is close to $t_{0}$, $H$ or some derivative of $H$ and therefore the curvature become singular.

Such choice of Hubble parameter corresponds to accelerated universe, becouse if Eq.(\ref{Hsingular}) is a solution of the EOM (\ref{GutenTag}), it is easy to see that the strong energy condition ($\rho_{\mathrm{eff}}+3p_{\mathrm{eff}}\geq 0$) is always violated when $\beta>0$, or is violated for small value of $t$ when $\beta<0$. It means that in any case the singularity could emerge as final evolution of accelerated universe.

The finite-time future singularities can be classified
in the following way\cite{Nojiri:2005sx}:
\begin{itemize}
\item Type I and Big Rip. 
It corresponds to $\beta>1$ and $\beta=1$. $H$ and $R$ ($\sim H^{2}$) diverge.
\item Type II (sudden\cite{altrosusing}).
It corresponds to $-1<\beta<0$. $R$ ($\sim \dot H$) diverges.
\item Type III. 
It corresponds to $0<\beta<1$. $H$ and $R$ ($\sim \dot H$) diverge.
\item Type IV.
It corresponds to
$\beta<-1$ but $\beta$ is not any integer number. Some derivative of $H$ and therefore the curvature becomes singular.
\end{itemize}
We note that in the present paper, we call singularities for $\beta=1$ and
those for $\beta>1$ as the ``Big Rip'' singularities and the ``Type I''
singularities, respectively.

\section{Reconstruction method}
\label{sec:4}

In order to study the finite-time singularities in $\mathcal{F}(R,G)$-gravity, we use the reconstruction method\cite{global}. 

We assume that the contribute of ordinary matter and radiation in expanding singular universe is too small with respect to the modified gravity, and we study the pure gravitational action of $\mathcal{F}(R,G)$-gravity,
i.e., the action in Eq.~(\ref{azione}) without
${\mathcal{L}}_{\mathrm{matter}}$.
In this case, it follows from Eqs.~(\ref{eq:2.7}) and (\ref{eq:2.8}) that
the EOM of $\mathcal{F}(R,G)$-gravity are given by\cite{Zerbini}:
\begin{eqnarray}
&&
24H^{3}{\dot{\mathcal{F}}}'_{G}
+6H^{2}{\mathcal{F}}'_{R}
+6H{\dot{\mathcal{F}}}'_{R}+(\mathcal{F}-R{\mathcal{F}}'_{R}
-G{\mathcal{F}}'_{G})=0\,,
\label{un} \\
&&
8H^{2}{\ddot{\mathcal{F}}}'_{G}+2{\ddot{\mathcal{F}}}'_{R}
+4H{\dot{\mathcal{F}}}'_{R}+16H{\dot{\mathcal{F}}}'_{G}(\dot H+H^{2})
\nonumber \\
&& \hspace{14mm}
+{\mathcal{F}}'_{R}(4\dot H+6H^{2})
+\mathcal{F}-R {\mathcal{F}}'_{R}
-G {\mathcal{F}}'_{G}=0\,.
\label{dos}
\end{eqnarray}
In the case of pure gravity,
these two equations are linearly dependents.

Moreover, we have
\begin{equation}
R = 6 \left(2H^{2}+\dot H \right)\,,
\quad
G = 24H^{2} \left( H^{2}+\dot H \right)\,.
\label{eq:G}
\end{equation}
It is easy to see that, in the case of Type I, II and III singularities, $G$ and $R$ tend to infinitive when $t\rightarrow t_{0}$ in Eq.(\ref{Hsingular}), and in the case of Type IV singularities tend to zero. 

By using proper functions $Z(t)$, $P(t)$ and $Q(t)$ of a
scalar field which is identified with the cosmic time $t$,
we can rewrite the action in Eq.(\ref{azione}) without
${\mathcal{L}}_{\mathrm{matter}}$ to
\begin{equation}
S=\frac{1}{2\kappa^2}\int d^{4}x \sqrt{-g} \left( Z(t)R+P(t)G+Q(t)
\right)\,.
\label{azionemodificata}
\end{equation}
By the variation with respect to $t$, we obtain
\begin{equation}
Z'(t)R+P'(t)G+Q'(t)=0\,, 
\label{t}
\end{equation}
from which in principle it is possible to find $t=t(R,G)$.
Here, the prime denotes differentiation with respect to $t$.
By substituting $t=t(R,G)$ into Eq.~(\ref{azionemodificata}),
we find the action in terms of $\mathcal{F}(R,G)$
\begin{equation}
\mathcal{F}(R,G)=Z(R,G)R+P(R,G)G+Q(R,G)\label{F(R,G)}\,.
\end{equation}

We describe the scale factor as
\begin{equation}
a(t)=a_{0}\exp(g(t))\,, \label{a}
\end{equation}
where $a_{0}$ is a constant and $g(t)$ is a function of time.
By using the Eqs.(\ref{un}), (\ref{dos}) and (\ref{a}), the matter conservation law $\dot\rho+3(\rho+p)$ and then neglecting the contribution from matter, we get the differential
equation
\begin{equation}
Z''(t)+4 \dot{g}^{2}(t)P''(t)-\dot g(t)Z'(t)+(8 \dot{g} \ddot{g}
-4 \dot{g}^{3}(t))P'(t)+2 \ddot{g}(t)Z(t)=0\,,
\label{Pi}
\end{equation}
where the Hubble parameter is $H(t)=\dot{g}(t)$.
By using Eq.~(\ref{un}), $Q(t)$ becomes
\begin{equation}
Q(t)=-24\dot{g}^{3}(t)P'(t)-6\dot{g}^{2}(t)Z(t)-6\dot{g}(t)Z'(t)\,.
\label{Qu}
\end{equation}
It means that in principle, by solving Eq.(\ref{Pi}) on the singular solution (\ref{Hsingular}), it is possible to reconstruct $\mathcal{F}(R,G)$ producing finite-time future singularities.\\ 
\\
In general, if $Z(t)\neq 0$, $\mathcal{F}(R,G)$ can be written in the
following form:
\begin{equation}
 \mathcal{F}(R,G)=R g(R,G)+f(R,G)\,,
\label{zuzzurellone}
\end{equation}
where $g(R,G)\neq 0$ and $f(R,G)$ are generic functions of $R$ and $G$. From
Eqs.~(\ref{un}) and (\ref{dos}), we obtain
\begin{eqnarray}
\rho_{\mathrm{eff}} &=& -\frac{1}{2\kappa^{2}g(R,G)}\biggl[
24H^{3}{\dot{\mathcal{F}}}'_{G}
+6H^{2}\left( R\frac{d g(R,G)}{d R}+\frac{d f(R,G)}{d R} \right)
+6H{\dot{\mathcal{F}}}'_{R}
\nonumber \\
&& \hspace{25mm}
+(\mathcal{F}-R{\mathcal{F}}'_{R}
-G{\mathcal{F}}'_{G})\biggr]\,,
\label{tic}\\
p_{\mathrm{eff}} &=& \frac{1}{2\kappa^{2}g(R,G)}\biggl[
8H^{2}{\ddot{\mathcal{F}}}'_{G}+2{\ddot{\mathcal{F}}}'_{R}
+4H{\dot{\mathcal{F}}}'_{R}+16H{\dot{\mathcal{F}}}'_{G}(\dot H+H^{2})
\nonumber \\
& & +\left(R\frac{d g(R,G)}{d R}+\frac{d f(R,G)}{d R}\right)(4\dot H+6H^{2})
+
\mathcal{F}-R {\mathcal{F}}'_{R}-G {\mathcal{F}}'_{G}\biggr]\,,
\label{tac}
\end{eqnarray}
where $\rho_{\mathrm{eff}}$ and $p_{\mathrm{eff}}$ are given by the
expressions in Eqs.(\ref{GutenTag}).

The modification of gravity could be included into the Equation of State (EoS) of an inhomogeneus dark fluid with energy density $\rho_{eff}$ and pressure $p_{eff}$
\begin{equation}
p_{eff}=\omega\rho_{eff}+\mathcal{G}(H,\dot{H}...)\,,\label{inEoS}
\end{equation}
where $\omega$ is the constant EoS parameter of matter and $\mathcal{G}(H,\dot{H}...)$ is a viscosity term given by
\begin{eqnarray}
\hspace{-5mm}
\mathcal{G}(H,\dot{H}...) &=&
\frac{1}{2\kappa^{2}g(R,G)} \biggl\{ (1+\omega)(\mathcal{F}
-R{\mathcal{F}}'_{R}-G{\mathcal{F}}'_{G})
\nonumber \\
\hspace{-5mm}
&&
+\left(R\frac{dg(R,G)}{dR}+\frac{df(R,G)}{dR}\right)
\left[6H^2(1+\omega)+4\dot{H}\right]
\nonumber \\
\hspace{-5mm}
& &
+ H{\dot{\mathcal{F}}}'_{R}(4+6\omega)
+8H{\dot{\mathcal{F}}}'_{G}\left[2\dot{H}+ H^{2}(2+3\omega)\right]
+2{\ddot{\mathcal{F}}}'_{R}+8 H^{2}{\ddot{\mathcal{F}}}'_{G} \biggr\}\,.
\label{Feldwebel}
\end{eqnarray}
The use of this
equation requires that $g(R,G)\neq 0$ on the solution.

By combining the two equations in Eq.(\ref{GutenTag}), we obtain
\begin{equation}
\mathcal{G}(H,\dot{H}...)=
-\frac{1}{\kappa^2}\left[ 2\dot{H}+3(1+\omega)H^2 \right]\,,
\label{prime}
\end{equation}
where Eq.(\ref{inEoS}) has been used.

\section{Singularities in $\mathcal{F}(R,G)$-gravity}
\label{sec:5}

\subsubsection*{Big Rip singularity in $F(G)$-gravity}

As a simple example of reconstruction method, we examine the Big Rip singularity in Gauss-Bonnet $F(G)$-gravity, where $\mathcal{F}(R,G)=R+F(G)$ and $F(G)$ is a function of Gauss-Bonnet invariant only. In this case, by putting $Z(t)=1$, the action of Eq.(\ref{azionemodificata}) can be written in terms of two proper functions $P(t)$ and $Q(t)$
and the variation with respect to $t$ yields
\begin{equation}
P'(t)G+Q'(t)=0\,, 
\label{t2}
\end{equation}
from which we can find $t=t(G)$ and the action in terms of $R$ and $F(G)$
\begin{equation}
F(G)=P(G)G+Q(G)\,. \label{F}
\end{equation}

Eq.(\ref{Pi}) and Eq.(\ref{Qu}) read
\begin{equation}
2 \frac{d}{d t} \left( \dot{g}^2(t)  \frac{d P(t)}{d t} \right) -2
\dot{g}^{3}(t) \frac{d P(t)}{d t} + \ddot{g}(t)=0\,,
\label{P}
\end{equation}
\begin{equation}
Q(t)= -24 \dot{g}^{3}(t)\frac{d P(t)}{d t}-6\dot{g}^2(t)\,.
\label{Q}
\end{equation}

For the Big Rip singularity, $\beta=1$ in Eq.~(\ref{Hsingular}). If we assume $H_{0}=0$ (the constant is negligible in the asyptotic singular limit $t\rightarrow t_{0}$), $\dot{g}(t)=h/(t_0-t)$ and 
the most general solution of Eq.~(\ref{P}) is given by
\begin{equation}
P(t)=\frac{1}{4h(h-1)}(2t_{0}-t)t+c_{1}\frac{(t_{0}-t)^{3-h}}{3-h}+c_{2}\,,
\end{equation}
where $c_{1}$ and $c_{2}$ are generic constants. From Eq.~(\ref{Q}),
we get
\begin{equation}
Q(t)=-\frac{6h^{2}}{(t_{0}-t)^2}-\frac{24 h^{3} \left[
\frac{(t_{0}-t)}{2h(h-1)}-c_{1}(t_{0}-t)^{2-h} \right]}{(t_{0}-t)^3}\,.
\end{equation}
Furthermore, from Eq.~(\ref{t2}) we obtain $t$ in terms of $G$ and,
by solving Eq.~(\ref{F}), we find
the most general form of $F(G)$ which realizes the Big Rip singularity
\begin{equation}
F(G)=\frac{\sqrt{6h^{3}(1+h)}}{h(1-h)}\sqrt{G}
+c_{1}G^{\frac{1+h}{4}}+c_{2}G\,.
\label{Garr}
\end{equation}
This is an exact solution of the EOM in the case of Big Rip.
The term $c_{2}G$ is a topological invariant.
In general, if for large values of $G$,
$F(G)\sim \alpha G^{1/2}$, where
$\alpha (\neq 0)$ is a constant,
the Big Rip singularity could appear for any value of $h\neq 1$.
Note that $c_{2} G^{(1+h)/4}$ is an invariant with
respect to the Big Rip solution.


\subsubsection*{Other types of singularities and more general $\mathcal{F}(R,G)$-gravity case}

In a similar way, it is possible to reconstruct $F(G)$-gravity models in wich the other types of singularities could appear, when $\beta\neq 1$ in Eq.(\ref{Hsingular}) and the scale factor, when $H_{0}=0$, behaves as
\begin{equation}
a(t)=\exp\left[\frac{h(t_0-t)^{1-\beta}}{\beta-1}\right]\,. \label{abis}
\end{equation}
We give some results.


The asymptotic solution (in the limit $t\rightarrow t_{0}$) of $F(G)$ when $\beta>1$ is expressed as
\begin{equation}
F(G)=-12\sqrt{\frac{G}{24}}\,.
\label{primo}
\end{equation}
Hence, if for large values of $G$, $F(G)\sim -\alpha\sqrt{G}$
with $\alpha>0$, a Type I singularity could appear.

When $\beta<1$, the asymptotic solution of $F(G)$
becomes
\begin{equation}
F(G)\sim \alpha |G|^{\gamma}\,,
\quad
\gamma = \frac{2\beta}{3\beta+1}\,,
\label{trentatre}
\end{equation}
where $\alpha$ is a constant. If  
for large values of $G$, $F(G)$ has this form with $0<\gamma <1/2$,
we find $0<\beta<1$ and a Type III singularity could emerge.
If for $G\rightarrow-\infty$,
$F(G)$ has the form in Eq.~(\ref{trentatre})
with $-\infty<\gamma<0$,
we find $-1/3<\beta<0$ and
a Type II (sudden) singularity could appear.
Moreover,
if for $G\rightarrow 0^{-}$,
$F(G)$ has the form in Eq.~(\ref{trentatre}) with
$1<\gamma<\infty$,
we obtain $-1<\beta<-1/3$ and a Type II singularity could occur.
If for $G\rightarrow 0^{-}$, $F(G)$ has the form in Eq.~(\ref{trentatre})
with $2/3<\gamma<1$,
we obtain $-\infty<\beta<-1$ and a Type IV singularity could appear.
We also require that $\gamma\neq2n/(3n-1)$,
where $n$ is a natural number.

As a consequence, a large class of realistic models of $F(G)$-gravity, which reproduce the current acceleration and the early-time
inflation, could generate future time-singularities, as for example\cite{Nojiri:2007bt}:
\begin{equation}
F_{1}(G) = \frac{a_{1}G^{n}+b_{1}}{a_{2}G^{n}+b_{2}}\,,\quad
F_{2}(G) = \frac{a_{1}G^{n+N}+b_{1}}{a_{2}G^{n}+b_{2}}\,,\quad
F_{3}(G) = a_{3} G^{n}(1+b_{3} G^{m})\,.
\end{equation}
All this models contain power functions of $G$ and for some choices of parameters could produce singularities.\\

With reconstruction method it is possible to derive also more general $\mathcal{F}(R,G)$-models producing finite-time singularities. For example, in the model $\mathcal{F}(R,G)=R-\alpha G/R$, where $\alpha$ is a positive constant, could appear the Type I singularity, whereas in the model $\mathcal{F}(R)=R+\alpha R^{\gamma}$, where $\alpha$ and $\gamma$ are constants, could appear Types II, III or IV singularities (for a review, see Ref.\cite{global}).

\section{Curing the finite-time future singularities}
\label{sec:6}

We discuss a possible way to cure the finite-time future
singularities in $F(G)$-gravity and $\mathcal{F}(R,G)$-gravity.
In the case of large curvature, the quantum effects become important
and lead to higher-order curvature corrections\cite{B1}.
It is therefore interesting to resolve the finite-time future singularities
with some power function of $G$ or $R$.

We consider the description of modified gravity as inhomogeneous fluid. If some singularities occur, Eq.~(\ref{prime}) behaves as
\begin{equation}
\mathcal{G}(H,\dot{H}...)\simeq
-\frac{3(1+\omega)h^{2}}{\kappa^2}(t_{0}-t)^{-2\beta}+\frac{2\beta h}{\kappa^{2}}(t_{0}-t)^{-\beta-1}\,. 
\label{casistiche}
\end{equation}
One way to prevent a singularity appearing could be
that the function $\mathcal{G}(H, \dot H...)$ becomes inconsistent with the
behavior of
Eq.~(\ref{casistiche}) in the singular limit ($t\rightarrow t_{0}$). 

Let us consider a simple example in order to cure Big Rip singularity in $F(G)$-gravity.
Suppose that for large values of $G$,
\begin{equation}
R+F(G\rightarrow \infty)\longrightarrow R+\gamma G^{m}\,,
\quad
m\neq 1
\,,
\label{eins}
\end{equation}
with $\gamma \neq 0$.
For
$H=h/(t_{0}-t)$, namely
the Big Rip case, we have
\begin{equation}
\mathcal{G}(H, \dot H...)\simeq \frac{\alpha}{(t_{0}-t)^{4m}}\,.
\end{equation}
Hence, if $m > 1/2$, $\mathcal{G}(H, \dot H...)$ tends to
infinity faster than Eq.~(\ref{casistiche}) and we avoid this kind of singularity.

As general results, we find that
the term $\gamma G^{m}$ with $m> 1/2$ and $m\neq 1$ cure the singularities
occurring when $G\rightarrow \pm\infty$ (Type I, II and III).
Moreover, the term $\gamma G^{m}$ with $m \leq 0$ cure the singularities occurring when $G\rightarrow 0^{-}$ (Type II, IV).

In $f(R)$-gravity (namely, $R$ plus a function of $R$),
by using the term $\gamma R^{m}$, the same consequences are found.
The term $\gamma R^{m}$ with $m>1$ cures the Type I, II and III
singularities. The term $\gamma R^{m}$ with $m<2$ cures the Type IV
singularity.

Within the framework of $\mathcal{F}(R,G)$-gravity, we can use the terms
such as $G^{m}/R^{n}$ to cure the singularities.
For example, we can avoid the Type I singularities if the asymptotic
behavior of the model is given by $\gamma G^{m}R^{n}$,  
with $m,n>0$.

\section{Effects of viscous fluid in singular universe}
\label{sec:7}

As the last point, we explore the role of perfect/viscous fluids within singular modified gravity, investigating
how the singularities may change or disappear, due to the contribution of quintessence or phantom fluids.

We consider the class of modified gravity $\mathcal{F}(R,G)=R+f(R,G)$, where $f(R,G)$ is a function of the Ricci scalar $R$ and the Gauss-Bonnet invariant $G$, and we suppose the presence in the universe of cosmic viscous fluid, whose EoS is given by 
\begin{equation}
p=\omega\rho-3 H\zeta\label{eq.state}\,,
\end{equation}
where $p$ and $\rho$ are the pressure and energy density of fluid, respectively, and $\omega$ is the EoS parameter. $\zeta$ is the bulk viscosity and in general it could depend on $\rho$, but we will consider the simplest case of constant viscosity only (for more general cases, see Ref.\cite{fluids}, \cite{Barrow}). On thermodynamical grounds, in order to have the positive sign of the entropy change in an irreversible process, $\zeta$ has to be a positive quantity.

The FRW-equations of motion are: 
\begin{equation}
\rho_{G}+\rho=\frac{3}{8\pi G_{N}}H^{2}\label{EOM1}\,,\quad
p_{G}+p=-\frac{1}{8 \pi G_{N}}\left(2\dot{H}+3H^2\right)\label{EOM2}\,.
\end{equation}
The modified gravity is formally included into the modified energy density $\rho_{G}$ and the modified pressure $p_{G}$, 
which correspond to Eqs.(\ref{tic})-(\ref{tac}) for $g(R,G)=1$.

The fluid energy conservation law is a consequence of the EOM (\ref{EOM1}):
\begin{equation}
\dot{\rho}+3H\rho(1+\omega)=9H^{2}\zeta\label{conservationlaw}\,.
\end{equation}

The presence of fluid could influence the behaviour of singular $f(R,G)$-models (i.e. models that in absence of 
fluids produce some singularities). We will check the solutions of the fluid energy density when $H$ is singular.
 

\subsubsection*{Non viscous case}
In the non-viscous case $\zeta=0$ (perfect fluid), the solution of Eq.(\ref{conservationlaw}) assumes the classical form:
\begin{equation}
\rho=\rho_{0}a(t)^{-3(1+\omega)}\,,\label{b}
\end{equation}
where $\rho_{0}$ is a positive constant and $a(t)$ is the scale factor of the universe. By combining Eq.(\ref{b}) with Eq.(\ref{abis}), it is easy to see that for $\beta>1$ (Type I singularity), $\rho$ grows up and diverges exponentially if $\omega<-1$. In the presence of phantom fluid, the EOM (\ref{EOM1}) become inconsistent with respect to the singular form of Hubble parameter in Eq.(\ref{Hsingular}), and the Type I singularity is not realized in $f(R,G)$-gravity.

When $0<\beta<1$, the fluid energy density $\rho$ is avoidable on the Type III singular solution, whereas
for Type II and IV singular models ($\beta<0$), the presence of quintessence or phantom fluids makes the singularities worse. In particular, in the case of $\beta<-1$, the dynamical behaviour of Eqs.(\ref{EOM1}) could become inconsistent, because $\rho$ behaves as $(t_{0}-t)$ and it is larger than the time-dependent part of $H^{2}$ ($\sim(t_{0}-t)^{-\beta}$) when $t\rightarrow t_{0}$. 

\subsubsection*{Constant viscosity}

Suppose to have the bulk viscosity equal to a constant, $\zeta=\zeta_{0}$, and the Hubble parameter in the general form of Eq.(\ref{Hsingular}).
The asymptotic solutions of Eq.(\ref{conservationlaw}) in the singular limit $t\rightarrow t_{0}$ are:
\begin{equation}
\rho\simeq \frac{3 h \zeta_{0}}{(1+\omega)(t_{0}-t)^{\beta}}\,,\phantom{spac}\beta>1\,,\label{zagzag}
\end{equation} 
\begin{equation}
\rho\simeq \frac{9\zeta_{0}h^{2}}{(2\beta-1)(t_{0}-t)^{2\beta-1}}\,,\phantom{space}1>\beta>0\,,\label{otto}
\end{equation}
\begin{equation}
\rho\simeq\frac{9h H_{0}\zeta_{0}}{(\beta-1)(t_{0}-t)^{\beta-1}}+\frac{3H_{0}\zeta_{0}}{1+\omega}\,,\phantom{sp}0>\beta\,,H_{0}\neq 0\,.\label{zag}
\end{equation}
In the first and second cases ($\beta>0$), it is possible to see that $\rho$ diverges more slowly than $H^{2}$, so that viscous fluid does not influence the asymptotically behaviour of Types I and III singular models in Eqs.(\ref{EOM1}), due to the constant viscosity.
  
In the third case, we consider fluids that tend to a non-negligible energy density when $\beta<0$. It automatically leads to $H_{0}\neq 0$ in Eq.(\ref{Hsingular}) and $\rho$ behaves as in Eq.(\ref{zag}). Large bulk viscosity $\zeta_{0}$ becomes relevant in the EOM. Moreover, if $\omega<-1$, the effective energy density (namely, $\rho_{G}+\rho$) could be negative and avoid the Type II and IV singularities for expanding universe (where $H_{0}>0$).

\section{Conclusion}
\label{sec:8}

We have investigated the
finite-time future singularities in $F(G)$-gravity and
$\mathcal{F}(R,G)$-gravity.
We can reconstruct the $F(G)$-gravity and
$\mathcal{F}(R,G)$-gravity models in which
the singularities could appear. Note that all types of future-time singularities could appear in $\mathcal{F}(R,G)$-modified gravity.
In addition, we have discussed a possible way to resolve
the finite-time future singularities in $F(G)$-gravity and
$\mathcal{F}(R,G)$-gravity under quantum effects of higher-order
curvature corrections or the presence of perfect/viscous fluid in the universe.
\begin{acknowledgement}
I wish to thank K. Bamba, S. D. Odintsov and S. Zerbini, whom the paper `Finite-time future singularities in modified Gauss-Bonnet and $\mathcal{F}(R,G)$ gravity and singularity avoidance' published in Eur. Phys. J. C67 has been written with.

\end{acknowledgement}

\end{document}